\documentclass[aps,prl,preprint,tightenlines,superscriptaddress,showpacs,byrevtex]{revtex4}

\usepackage{graphicx}
\usepackage{epsfig}
\usepackage{dcolumn}

\newcommand{\yones}{\Upsilon(1S)}
\newcommand{\x}{X(3872)}

\newcommand{\cdfy}{Y(4140)}

\newcommand{\eff}{\varepsilon}
\newcommand{\BR}{{\cal B}}

\newcommand{\piz}{\pi^0}

\newcommand{\etac}{\eta_c}
\newcommand{\chicz}{\chi_{c0}}
\newcommand{\chico}{\chi_{c1}}
\newcommand{\chict}{\chi_{c2}}
\newcommand{\chicJ}{\chi_{cJ}}

\newcommand{\psp}{\psi(2S)}

\newcommand{\jpsi}{J/\psi}

\newcommand{\EE}{e^+e^-}
\newcommand{\MM}{\mu^+\mu^-}
\newcommand{\LL}{\ell^+\ell^-}
\newcommand{\pp}{\pi^+\pi^-}
\newcommand{\kk}{K^+K^-}

\newcommand{\kkpi}{K_S^0K^+\pi^- + c.c.}
\newcommand{\ppjpsi}{\pi^+\pi^- J/\psi}

\newcommand{\beq}{\begin{equation}}
\newcommand{\eeq}{\end{equation}}
\newcommand{\bitm}{\begin{itemize}}
\newcommand{\eitm}{\end{itemize}}

\def\Journal#1#2#3#4{{#1} {\bf #2}, #3 (#4)}

\def\PRL{Phys. Rev. Lett.}

\begin{document}

\preprint{} \preprint{\vbox{ \hbox{   }
                        \hbox{Belle Preprint 2010-15}
                        \hbox{KEK   Preprint 2010-24}
                        \hbox{BIHEP-EP-2010-001}
                        }}
\title{
\quad\\[1.0cm]
Search for charmonium and charmonium-like states in $\Upsilon(1S)$
radiative decays}

\affiliation{Budker Institute of Nuclear Physics, Novosibirsk}
\affiliation{Faculty of Mathematics and Physics, Charles University,
Prague}
\affiliation{Chiba University, Chiba}
\affiliation{Department of Physics, Fu Jen Catholic University,
Taipei}
\affiliation{Justus-Liebig-Universit\"at Gie\ss{}en,
Gie\ss{}en}
\affiliation{The Graduate University for Advanced
Studies, Hayama}
\affiliation{Hanyang University, Seoul}
\affiliation{University of
Hawaii, Honolulu, Hawaii 96822}
\affiliation{High Energy Accelerator
Research Organization (KEK), Tsukuba}
\affiliation{Institute of High Energy Physics, Chinese Academy of
Sciences, Beijing}
\affiliation{Institute of High Energy Physics,
Vienna}
\affiliation{Institute of High Energy Physics, Protvino}
\affiliation{INFN - Sezione di Torino, Torino}
\affiliation{Institute for Theoretical and Experimental Physics,
Moscow}
\affiliation{J. Stefan Institute, Ljubljana}
\affiliation{Kanagawa University, Yokohama}
\affiliation{Institut f\"ur Experimentelle Kernphysik, Karlsruher
Institut f\"ur Technologie, Karlsruhe}
\affiliation{Korea University, Seoul}
\affiliation{Kyungpook National University, Taegu}
\affiliation{\'Ecole Polytechnique F\'ed\'erale de Lausanne (EPFL),
Lausanne} \affiliation{Faculty of Mathematics and Physics,
University of Ljubljana, Ljubljana}
\affiliation{University of
Maribor, Maribor} \affiliation{Max-Planck-Institut f\"ur Physik,
M\"unchen} \affiliation{University of Melbourne, School of Physics,
Victoria 3010} \affiliation{Nagoya University, Nagoya}
\affiliation{Nara Women's University, Nara}
\affiliation{National United University, Miao Li}
\affiliation{Department of Physics, National Taiwan University,
Taipei}
\affiliation{H. Niewodniczanski Institute of Nuclear
Physics, Krakow}
\affiliation{Nippon Dental University, Niigata}
\affiliation{Niigata University, Niigata}
\affiliation{University of
Nova Gorica, Nova Gorica}
\affiliation{Novosibirsk State University,
Novosibirsk}
\affiliation{Osaka City University, Osaka}
\affiliation{Panjab University, Chandigarh}
\affiliation{University of Science and Technology of China, Hefei}
\affiliation{Seoul National University, Seoul}
\affiliation{Sungkyunkwan University, Suwon}
\affiliation{School of
Physics, University of Sydney, NSW 2006}
\affiliation{Tata Institute
of Fundamental Research, Mumbai}
\affiliation{Excellence Cluster
Universe, Technische Universit\"at M\"unchen, Garching}
\affiliation{Tohoku Gakuin University, Tagajo}
\affiliation{Tohoku
University, Sendai}
\affiliation{Department of Physics, University
of Tokyo, Tokyo}
\affiliation{IPNAS, Virginia Polytechnic Institute and State
University, Blacksburg, Virginia 24061}
\affiliation{Wayne State
University, Detroit, Michigan 48202}
\affiliation{Yonsei University,
Seoul}
  \author{C.~P.~Shen}\affiliation{Institute of High Energy Physics, Chinese Academy of Sciences, Beijing}\affiliation{University of Hawaii, Honolulu, Hawaii 96822} 
  \author{X.~L.~Wang}\affiliation{Institute of High Energy Physics, Chinese Academy of Sciences, Beijing} 
  \author{C.~Z.~Yuan}\affiliation{Institute of High Energy Physics, Chinese Academy of Sciences, Beijing} 
  \author{P.~Wang}\affiliation{Institute of High Energy Physics, Chinese Academy of Sciences, Beijing} 
  \author{I.~Adachi}\affiliation{High Energy Accelerator Research Organization (KEK), Tsukuba} 
  \author{H.~Aihara}\affiliation{Department of Physics, University of Tokyo, Tokyo} 
  \author{T.~Aushev}\affiliation{\'Ecole Polytechnique F\'ed\'erale de Lausanne (EPFL), Lausanne}\affiliation{Institute for Theoretical and Experimental Physics, Moscow} 
  \author{A.~M.~Bakich}\affiliation{School of Physics, University of Sydney, NSW 2006} 
  \author{V.~Balagura}\affiliation{Institute for Theoretical and Experimental Physics, Moscow} 
  \author{K.~Belous}\affiliation{Institute of High Energy Physics, Protvino} 
% ...
\collaboration{The Belle Collaboration}


\date{\today}

\begin{abstract}

Using a sample of 102 million $\Upsilon(1S)$ events collected with
the Belle detector, we report on the first search for
charge-parity-even charmonium and charmonium-like states in
$\Upsilon(1S)$ radiative decays. No significant $\chi_{cJ}$ or
$\eta_c$ signal is observed and 90\% C.L. limits on
$\BR(\Upsilon(1S)\to \gamma \chi_{c0})<6.5 \times 10^{-4}$,
$\BR(\Upsilon(1S)\to \gamma \chi_{c1})<2.3\times 10^{-5}$,
$\BR(\Upsilon(1S)\to \gamma \chi_{c2})<7.6 \times 10^{-6}$, and
$\BR(\Upsilon(1S)\to \gamma \eta_c)<5.7\times 10^{-5}$ are obtained.
The product branching fraction limits $\BR(\Upsilon(1S)\to \gamma
X(3872)) \BR(X(3872)\to\pi^+\pi^-\jpsi)< 1.6 \times 10^{-6}$,
$\BR(\Upsilon(1S)\to \gamma X(3872)) \BR(X(3872)\to\pi^+\pi^-\pi^0
\jpsi)< 2.8\times 10^{-6}$, $\BR(\Upsilon(1S)\to \gamma X(3915))
\BR(X(3915)\to\omega \jpsi)< 3.0\times 10^{-6}$, and
$\BR(\Upsilon(1S)\to \gamma Y(4140)) \BR(Y(4140)\to \phi
\jpsi)<2.2\times 10^{-6}$ are obtained at the 90\% C.L. Furthermore,
no evidence is found for excited charmonium states below
4.8~GeV/$c^2$.

\end{abstract}

\pacs{14.40.Nd, 14.20.Lq, 13.25.Gv}

\maketitle

There is renewed interest in charmonium spectroscopy after the
operation of the two $B$-factories. In addition to many conventional
charmonium states, a number of states with unusual properties have
been discovered, which may include states beyond the quark-model,
such as quark-gluon hybrids, meson molecules, multi-quark states,
and so
on~\cite{bellex,babary,belley,babar_pppsp,belle_pppsp,belle_z,cdfy}.
States with $J^{PC}=1^{--}$ can be studied using initial state
radiation (ISR) in the large $\Upsilon(4S)$ data samples. For the
study of charge-parity-even charmonium states, radiative decays of
the $\Upsilon$ states below open-bottom threshold are used.

The production rates of the lowest lying $P$-wave spin-triplet
($\chicJ$, $J$=0, 1, or 2) and $S$-wave spin-singlet ($\etac$)
states in $\yones$ radiative decays are calculated in
Ref.~\cite{ktchao}, where the former is at the part per million
level, and the latter is about $5\times 10^{-5}$. There are no
calculations for radiative decays involving  excited charmonium
states, let alone for charmonium-like states, such as the
$\x$~\cite{bellex}, the $X(3915)$~\cite{uehara}, and the
$\cdfy$~\cite{cdfy}.

In this paper, we report on a search for the $\chicJ$, $\etac$,
$\x$, $X(3915)$, and $\cdfy$ states in $\yones$ radiative decays.
The $\chicJ$ states are reconstructed via their $E1$ transition to
the $\jpsi$. The $\etac$ is reconstructed in the $\kkpi$, $\pp\kk$,
$2(\kk)$, $2(\pp)$, and $3(\pp)$ final states. To search for the
$\x$ and $X(3915)$, we use the $\ppjpsi$ and $\pp\piz\jpsi$ final
states, while we reconstruct the $\cdfy$ in the $\phi\jpsi$ mode.
This analysis is based on a 5.7~fb$^{-1}$ data sample collected at
the $\Upsilon(1S)$ (102 million $\yones$ events) and a 1.8~fb$^{-1}$
data sample collected at $\sqrt{s}=9.43$~GeV (continuum data) with
the Belle detector~\cite{Belle} operating at the KEKB
asymmetric-energy $\EE$ collider~\cite{KEKB}.

For each charged track, the impact parameters perpendicular to and
along the beam direction with respect to the interaction point are
required to be less than 0.5~cm and 4~cm, respectively, and the
transverse momentum must exceed 0.1~GeV/$c$ in the laboratory frame.
For each charged track, information from different detector
subsystems is combined to form a likelihood $\mathcal{L}_i$ for each
particle species~\cite{pid}. Tracks with
$\mathcal{R}_K=\frac{\mathcal{L}_K}{\mathcal{L}_K+\mathcal{L}_\pi}>0.6$
are identified as kaons, tracks with $\mathcal{R}_K<0.4$ are
identified as pions. With these selections, the kaon (pion)
identification efficiency is about 90\% (96\%), while 5\% (6\%) of
kaons (pions) are misidentified as pions (kaons). For electron
identification, the likelihood ratio is defined as
$\mathcal{R}_e=\frac{\mathcal{L}_e}{\mathcal{L}_e+\mathcal{L}_x}$,
where $\mathcal{L}_e$ and $\mathcal{L}_x$ are the likelihoods for
electron and non-electron hypotheses, respectively. These are
determined using the ratio of the energy deposited in the
electromagnetic calorimeter (ECL) to the momentum measured in the
silicon vertex detector (SVD) and central drift chamber (CDC), the
shower shape in the ECL, the matching between the position of the
charged track trajectory and the cluster position in the ECL, hit
information from the aerogel threshold Cherenkov counters (ACC), and
dE/dx information in the CDC~\cite{EID}. For muon identification,
the likelihood ratio is defined as
$\mathcal{R}_\mu=\frac{\mathcal{L}_\mu}{\mathcal{L}_\mu+\mathcal{L}_\pi+\mathcal{L}_K}$,
where $\mathcal{L}_\mu$, $\mathcal{L}_\pi$ and $\mathcal{L}_K$ are
the likelihoods for muon, pion and kaon  hypotheses, respectively.
These are based on track matching quality and penetration depth of
associated hits in the iron flux-return (KLM)~\cite{MUID}.

We reconstruct $\jpsi$ mesons from $\EE$ or $\MM$ candidates. In
order to reduce the effect of bremsstrahlung or final-state
radiation, photons detected in the ECL within 0.05~radians of the
original $e^+$ or $e^-$ direction are included in the calculation of
the $\EE (\gamma)$ invariant mass. For electrons from $\jpsi \to
\EE$, one track should have $\mathcal{R}_e>0.95$ and the other
$\mathcal{R}_e>0.05$; for muons from $\jpsi \to \mu^+ \mu^-$, at
least one track should have $\mathcal{R}_\mu>0.95$ (in the $\chicJ$
analysis, the other track should have associated hits in the KLM
detector that agree with the extrapolated trajectory of a charged
track provided by the drift chamber). The lepton identification
efficiency is about 90\% for $\jpsi \to e^+ e^-$ and 87\% for $\jpsi
\to \mu^+ \mu^-$. In order to improve the $\jpsi$ momentum
resolution, a mass fit to the reconstructed $\jpsi$ candidates is
then performed for all the channels with $\jpsi$ signals.

A neutral cluster is considered to be a photon candidate if its ECL
shower does not match the extrapolation of any charged track and the
energy deposition is greater than 40~MeV. The photon candidate with
the maximum energy in the $\EE$ center-of-mass (C.M.) frame is taken
to be the $\yones$ radiative decay photon, and its energy is
required to be greater than $3.5~\hbox{GeV}$, which corresponds to a
4.8~GeV/$c^2$ mass particle produced in $\Upsilon(1S)$ radiative
decays.


To study the $\gamma\chicJ$ mode, we reconstruct $\chicJ$ via its
decay into $\gamma\jpsi$. The deposited energy of $\chicJ$'s photon
is required to be greater than 150~MeV, and the total number of
photons in the event is required to be exactly two, in order to
suppress multi-photon backgrounds. The higher energy photon is
denoted as $\gamma_h$ and the lower energy one is denoted as
$\gamma_l$. The angle between the two photons should be larger than
$18^{\circ}$ to remove the background from split-off fake photons.
To remove the ISR background $\EE\to \gamma_{\rm ISR}\psp\to
\gamma_{\rm ISR} \gamma \chicJ$, where a photon is missed, we
require the square of the missing mass of $\gamma_l$ and lepton-pair
to be within $-0.5~{\rm GeV}^2/c^4$ and 0.5~GeV$^2/c^4$ since this
background has at least two missing photons (the $\gamma_{{\rm
ISR}}$ photon(s) and one photon from the $\psp$ decay) and the
missing mass tends to be large. Bhabha and dimuon background events
with final-state radiative photons are further suppressed by
removing events where a photon is detected within a $18^\circ$ cone
around each charged track direction.

A clear $\jpsi$ signal is observed in the $\MM$ mode, while no
significant $\jpsi$ signal is observed in the $\EE$ mode due to
residual radiative Bhabha background. The $\jpsi$ signal region is
defined as $|m_{\ell^+\ell^-}-m_{\jpsi}|<30~\hbox{MeV}/c^2$
($\approx 2.5\sigma$), and the $\jpsi$ mass sidebands are defined as
$2.959~\hbox{GeV}/c^2<m_{\ell^+\ell^-}<3.019~\hbox{GeV}/c^2$ or
$3.175~\hbox{GeV}/c^2<m_{\ell^+\ell^-}<3.235~\hbox{GeV}/c^2$; where
the latter is twice as wide as the signal region.

\begin{figure}[htbp]
\psfig{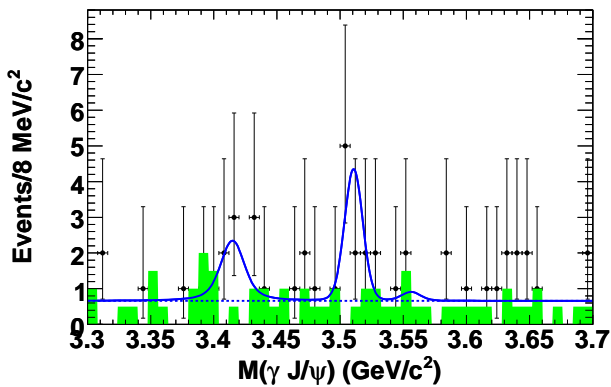} \caption{ The
$\gamma_{l}\jpsi$ invariant mass distribution in the $\yones$ data
sample.  Hints of $\chi_{c0}$ and $\chi_{c1}$ signals are seen
although no obvious $\chict$ signal is observed. The solid curve is
the best fit, the dashed curve is the background, and the shaded
histogram is from the normalized $\jpsi$ mass sidebands.}
\label{mgll}
\end{figure}

Figure~\ref{mgll} shows the $\gamma_l\jpsi$ invariant mass
distribution after the above selections are applied to the $\yones$
data sample for the combined $\EE$ and $\MM$ modes, together with
the background estimated from the normalized $\jpsi$ mass sidebands.
Apart from possible weak $\chi_{c0}$ and $\chi_{c1}$ signals, the
$\jpsi$ sideband events represent well the signal region, indicating
that the production of any of the $\chicJ$ states is  not
significant. There are no structures at higher masses, where we
would expect excited $\chicJ$ states.

A fit to the signal region is performed with Breit-Wigner (BW)
functions for the resonances convolved with Gaussian resolution
functions and a second-order polynomial background term.  This fit
yields $5.9^{+3.9}_{-3.1}$, $8.5^{+3.8}_{-3.1}$, and
$0.6^{+2.1}_{-1.4}$ events for the $\chicz$, $\chico$, and $\chict$,
respectively. Here the width of the Gaussian resolution function is
fixed as $7.0$~MeV/$c^2$, its MC-determined value. Bayesian upper
limits on the number of events at the 90\% C.L. by integrating the
likelihood distribution (as a function of the yield) are found to be
$11.5$, $13.8$, and $2.4$ for the $\chicz$, $\chico$, and $\chict$,
respectively.


To study the $\gamma\etac$ mode, we reconstruct the $\etac$ mass
from the invariant masses of $\kkpi$, $\pp\kk$, $2(\kk)$, $2(\pp)$,
and $3(\pp)$. Well-measured charged tracks are selected and the
numbers of charged tracks are six for the $3(\pp)$ final state and
four for the other final states. All the charged tracks are required
to be identified as kaons or pions. The recoil mass-squared of the
charged particles in each $\etac$ decay mode is required to be
within $-1$~GeV$^2/c^4$ and 1~GeV$^2/c^4$. For $K_S^0$ candidates
decaying into $\pp$ in the $\kkpi$ mode, we require that the
invariant mass of the $\pp$ pair lie within 30 MeV/$c^2$ of the
$K^0_S$ nominal mass and that the $K^0_S$ candidate must have a
displaced vertex and flight direction consistent with a $K^0_S$
originating from the IP; the same selection method is used in
Ref.~\cite{ks}. There are events with leptons misidentified as pions
in the $\pp \kk$ and $2(\pp)$ modes, and they are removed by
requiring $\mathcal{R}_e<0.9$ and $\mathcal{R}_\mu<0.9$ for the pion
candidates.

\begin{figure}[htb]
\centerline{\psfig{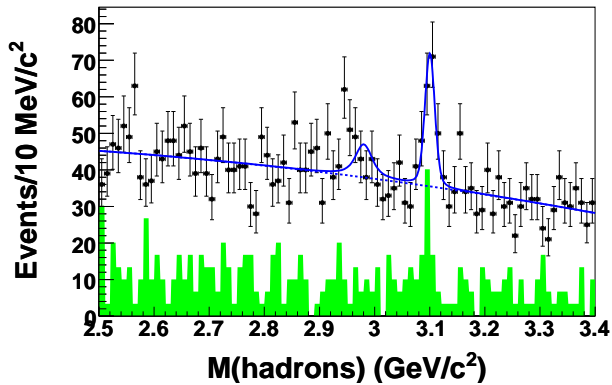}} \caption{
The mass distribution for a sum of the five $\etac$ decay modes. The
solid line is a sum of the corresponding functions obtained from a
simultaneous fit to all the $\etac$ decay modes, and the dashed line
is a sum of the background functions from the fit. The shaded
histogram is a sum of the normalized continuum events, where the
$\jpsi$ signal is produced via ISR.} \label{metac}
\end{figure}

Figure~\ref{metac} shows the combined mass distribution for the five
$\etac$ decay modes after the selection described above. The peak in
hadronic mass at the $\jpsi$ mass, as seen in Fig.~\ref{metac}, can
be attributed to the ISR process, $\EE \to \gamma_{{\rm ISR}}
\jpsi$, while the accumulation of events within the $\etac$ mass
region is small. The shaded histogram in Fig.~\ref{metac} is the
same distribution for the continuum data, normalized according to
the ratio of the luminosities on and off the $\yones$ peak. From
Fig.~\ref{metac}, we can see that the $\jpsi$ signal in $\yones$
data is well reproduced by the normalized continuum data,
demonstrating its ISR origin. It is also evident that $\yones$
radiative decays to light hadrons are substantial, as indicated by
the difference between the number of non-resonant events in the two
data sets.

We perform a simultaneous fit to all the $\etac$ decay modes, where
the $\etac$ mass and width are taken from the PDG~\cite{PDG}, and
the ratio of the yields in all the channels is fixed to
$\BR_i\epsilon_i$, where each $\BR_i$ is the $\etac$ decay branching
fraction for the $i$-th mode reported by the PDG~\cite{PDG}, and
$\epsilon_i$ is the MC-determined efficiency for this mode. In the
fit, we take a BW function convolved with a Gaussian resolution
function (its resolution is fixed to 7.9~$\hbox{MeV}/c^2$ from MC
simulation) as the $\etac$ signal shape, another Gaussian function
as the $\jpsi$ signal shape, and a second-order polynomial as the
background shape. The fitted results are shown in Fig.~\ref{metac},
where the solid line is the sum of the best fit functions in the
simultaneous fit, and the dashed line is the sum of the background
functions. The fit yields $46\pm 22$ $\etac$ signal events, with a
statistical significance of $2.2\sigma$. An upper limit on the
number of the $\etac$ signal events is estimated to be 72 at the
90\% C.L. From the fit, we obtain $89\pm 20$ and $54\pm16$ $\jpsi$
signal events in $\yones$ and normalized continuum data samples,
respectively, with a mass of $3099.9\pm 2.1~\hbox{MeV}/c^2$, which
is consistent with PDG value.

\begin{figure}[htbp]
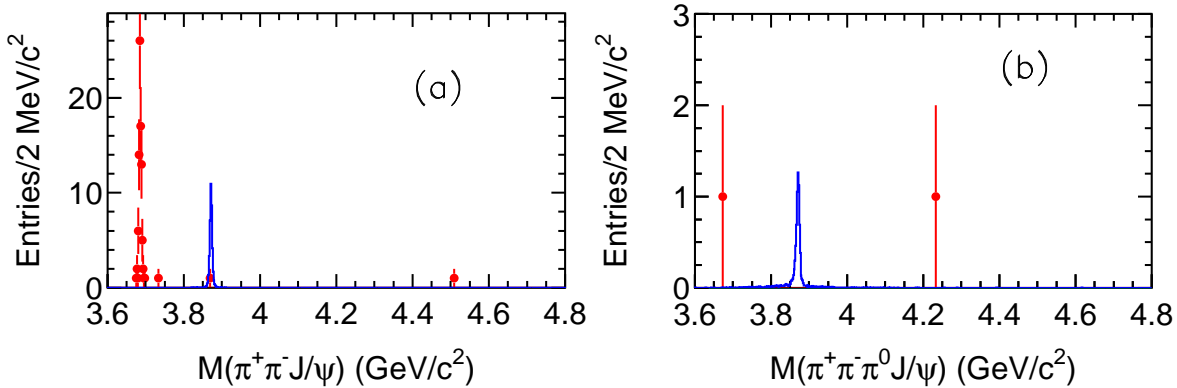

 \psfig{file=fig3a.epsi, height=5cm, angle=0}
 \psfig{file=fig3b.epsi, height=5.1cm, angle=0}
\caption{(a) Distribution of the $\ppjpsi$ invariant mass for
$\yones\to \gamma\ppjpsi$ candidates. (b) Distribution of the
$\pp\piz\jpsi$ invariant mass for $\yones\to \gamma\pp\piz\jpsi$
candidates. Points with error bars are data, open histograms are the
MC expectation for the $X(3872)$ signal (not normalized). The peak
at $3.686$~GeV/$c^2$ in (a) is due to $\psp$ production via ISR.}
\label{x3872}
\end{figure}

The selection criteria for $\yones\to \gamma\x$, $\x\to \ppjpsi$ are
similar to those used for ISR $\ppjpsi$ events in $\Upsilon(4S)$
data~\cite{belley}.  We require that one $\jpsi$ candidate be
reconstructed, that two well identified $\pi$'s have an invariant
mass greater than 0.35~GeV/$c^2$, and that the recoil mass-squared
of the $\ppjpsi$ be between $-1$~GeV$^2/c^4$ and 1~GeV$^2/c^4$. To
suppress ISR $\ppjpsi$ background, we require that the polar angle
of the radiative photon satisfy $|\cos\theta|<0.9$ in the $\EE$ C.M.
system. Except for a few remaining ISR produced $\psp$ signal
events, only a small number of events appear above the $\psp$ peak
in the $\pp\jpsi$ invariant mass distribution, as shown in
Fig.~\ref{x3872}(a). Within the $X(3872)$ signal region, there is
one event with a mass of 3.870 GeV/$c^2$. However, there are no
events in the $\jpsi$ mass sidebands from 3.6 to 4.8 GeV/$c^2$. We
estimate the statistical significance of the $X(3872)$ signal to be
$2.3\sigma$ if the background distribution is flat above
3.7~GeV/$c^2$. Assuming that the number of signal events follows a
Poisson distribution with a uniform prior probability density
function and there is no background, the upper limit on the number
of the $X(3872)$ signal events is 3.9~\cite{PDG}.

We validate our analysis by measuring the $\psp$ ISR production
cross section as observed in the $\ppjpsi$ mode. By relaxing the
photon polar angle requirement, we observe 383 $\psp$ signal events
and a cross section of $\EE\to \gamma_{\rm ISR}\psp$ is measured to
be $(20.2\pm 1.1~(\hbox{stat.}))$~pb, in agreement with a
theoretical calculation of $18.5$~pb using PDG~\cite{PDG} values for
the $\psp$ resonance parameters as input.

To study the $\gamma \pp\piz\jpsi$ mode, we require the invariant
mass of a pair of photons to be within 10~MeV$/c^2$ around the
nominal $\piz$ mass (the mass resolution is about 4~MeV/$c^2$) to
select $\piz$ candidates. The other event selection criteria are
similar to those in the $X(3872) \to \ppjpsi$ mode, except that we
do not require the $\pp$ invariant mass to be greater than
0.35~GeV/$c^2$, which is used to remove the $\gamma$ conversion
background events in the $\ppjpsi$ mode.

Figure~\ref{x3872}(b) shows the $\pp\piz\jpsi$ invariant mass
distribution, where the open histogram is the MC expectation for the
$X(3872)$ signal shape plotted with an arbitrary normalization. We
observe two events in the $\pp \pi^0 \jpsi$ mass spectrum between
3.6~GeV/$c^2$ and 4.8~GeV/$c^2$ in the $\yones$ data. For these two
events, the $\pp \pi^0 \jpsi$ masses are 3.67~GeV/$c^2$ and
4.23~GeV/$c^2$, and the corresponding $\pp \pi^0$ masses are
0.54~GeV/$c^2$ and 1.04~GeV/$c^2$, respectively. The event at
3.67~GeV/$c^2$, is likely to be from $\EE \to \gamma_{\rm ISR} \eta
\jpsi \to \gamma_{\rm ISR} \pp \pi^0 \LL$, since 0.9 events are
expected from MC simulation. No event is observed within the
$X(3872)$ or $X(3915)$ mass region. An upper limit on the number of
$X(3872)$ or $X(3915)$ signal events is 2.3 at the 90\%
C.L.~\cite{PDG}.

We also search for the $Y(4140)$ in its decays into $\phi\jpsi$,
with $\phi\to \kk$ and $\jpsi\to \LL$. The selection criteria are
very similar to the analysis of $X(3872)\to \ppjpsi$ above. Here two
kaons are required to be positively identified and one $\jpsi$
candidate is reconstructed. No clear $\jpsi$ or $\phi$ signal can be
seen after the initial event selection. We define the $\jpsi$ signal
region as $|m_{\ell^+\ell^-}-m_{\jpsi}|<30~\hbox{MeV}/c^2$, and the
$\phi$ signal region as
$1.01~\hbox{GeV}/c^2<m_{\kk}<1.03$~GeV/$c^2$, according to MC
simulation. After applying all of the above event selection
criteria, there are no candidate events in the $\phi\jpsi$ invariant
mass region between 4~GeV/$c^2$ and 4.8~GeV/$c^2$.  An upper limit
on the number of $Y(4140)$ signal events is 2.3 at the 90\%
C.L.~\cite{PDG}.

There are several sources of systematic error in determining limits
on the branching fractions.  A particle identification efficiency
uncertainty between 2.4\%-3.7\% is assigned depending on the final
state particles.  An uncertainty in the tracking efficiency for
tracks with angles and momenta characteristic of signal events is
about 1\% per track, and is additive. Photon reconstruction
contributes an additional 2\% per photon.  Errors on the branching
fractions of the intermediate states are taken from the
PDG~\cite{PDG}. For the $\etac$ decays, the biggest difference in
the efficiency by using a phase space distribution and including
possible intermediate resonance states is 2.1\%. The difference in
overall efficiency for a flat radiative photon angular distribution
and a $1\pm \cos^2\theta$ distribution is less than 3.0\%.
Therefore, we quote an additional error of 5\% for all the states
studied due to limited knowledge of their decay dynamics. According
to MC simulation, the trigger efficiency is rather high, with an
uncertainty that is smaller than 1\%.  The uncertainty due to the
missing mass squared requirement is 1.0\% for the channels with only
one photon and 4.7\% for channels with more than one photon.
Uncertainties on the $\chi_{cJ}$ and $\etac$ signal event yields are
estimated to be 1.6\% and 15\%, respectively, by changing the order
of the background polynomial, the range of the fit, and the values
of the masses and widths of the resonances. In the $\yones \to
\gamma \chi_{cJ}$ mode, the uncertainty that is associated with the
requirement on the number of photons is 2\% after applying a
correction factor of 0.96 to the MC efficiency, which is determined
from a study of a very pure $\yones \to \mu^+ \mu^-$ event sample.
In the $\etac \to \kkpi$ mode, the uncertainty in the $K_S$
efficiency is determined by comparing yields for a sample of high
momentum $K_S \to \pi^+ \pi^-$ decays before and after applying the
$K_S$ candidate selection criteria; the efficiency difference
between data and MC simulation is less than 4.9\%~\cite{ks-error}.
Finally, the uncertainty on the total number of $\yones$ events is
2.2\%. Assuming that all of these systematic error sources are
independent, we add them in quadrature to obtain total systematic
errors as shown in Tab.~\ref{summary}.
In order to calculate conservative upper limits on these branching
fractions, the efficiencies have been lowered by a factor of
$1-\sigma_{\rm sys}$.

In summary, Table~\ref{summary} lists the final results for the
upper limits on the branching fractions of all the states studied,
together with the upper limits on the numbers of signal events and
their detection efficiencies.  The results obtained on the $\chicJ$
and $\etac$ production rates are not in contradiction with the
calculations in Ref.~\cite{ktchao}. No $\x$, $X(3915)$, or $\cdfy$
signals are observed, and the production rates of the $\ppjpsi$,
$\pp\piz\jpsi$, $\omega \jpsi$, or $\phi\jpsi$ modes are found to be
less than a few times $10^{-6}$ at the 90\% C.L. Furthermore, we
find no evidence for excited charmonium states below 4.8~GeV/$c^2$.

\begin{table}[htbp]
\caption{Summary of the limits on $\yones$ radiative decays to
charmonium and charmonium-like states $R$. $N^{\rm UP}_{\rm sig}$ is
the upper limit on the number of signal events, $\eff$ is the
efficiency,  $\sigma_{\rm sys}$ is the total systematic error and
$\BR(\yones \to \gamma R)^{\rm UP}$ (${\cal B}_R$) is the upper
limit at the 90\% C.L. on the decay branching fraction in the
charmonium state case, and on the product branching fraction in the
charmonium-like state case.} \label{summary}
\begin{center}
\begin{tabular}{c  c  c  c c}
\hline
 State ($R$) & $N^{\rm UP}_{\rm sig}$ &
$\eff$(\%) & $\sigma_{\rm
sys}$(\%)& ${\cal B}_R (10^{-5})$ \\
\hline
$\chi_{c0}$ & 11.5& 15.1 &11  & 65 \\
$\chi_{c1}$ & 13.8& 17.0 &11  & 2.3\\
$\chi_{c2}$ & 2.4& 15.8& 11   &0.76\\
$\etac$ & 72 & 25.1 & 23 &    5.7 \\
$\x \to \ppjpsi$ & 3.9 &23.2 &7.6 & 0.16 \\
$\x \to \pp \piz \jpsi$   &2.3 & 7.6  & 9.7 &0.28 \\
$X(3915) \to \omega \jpsi$  &2.3 & 8.1 & 9.7 &0.30 \\
$Y(4140) \to \phi \jpsi$ & 2.3 & 19.4 & 7.7 &0.22 \\
 \hline
\end{tabular}
\end{center}
\end{table}


We thank the KEKB group for excellent operation of the
accelerator, the KEK cryogenics group for efficient solenoid
operations, and the KEK computer group and the NII for valuable
computing and SINET3 network support. We acknowledge support from
MEXT, JSPS and Nagoya's TLPRC (Japan); ARC and DIISR (Australia);
NSFC (China); DST (India); MEST, KOSEF, KRF (Korea); MNiSW
(Poland); MES and RFAAE (Russia); ARRS (Slovenia); SNSF
(Switzerland); NSC and MOE (Taiwan); and DOE (USA).

\end{document}